\documentclass{article}

\usepackage{graphicx}
\usepackage{amsfonts}
\usepackage{amsmath}
\usepackage{amsbsy}
\usepackage{subfigure}
\usepackage{setspace} 
\usepackage{bm}
\usepackage{multicol}
\usepackage{multirow}

\begin{document}

\title{Magnetic Braiding and Parallel Electric Fields}
\author{A. L. Wilmot-Smith, G. Hornig  and D. I. Pontin.}
\maketitle

\begin{abstract}
\noindent

The braiding of the solar coronal magnetic field via photospheric motions 
-- with subsequent relaxation and 
magnetic reconnection --  is one of the most widely debated ideas of solar physics.
We readdress the theory in the light of developments in three-dimensional  magnetic
reconnection theory.  It is known that the  integrated parallel electric field 
along field lines is the key quantity determining the rate of reconnection, 
in contrast with the two-dimensional case
where the electric field itself is the important quantity.
We demonstrate that this difference becomes crucial for sufficiently complex
magnetic field structures.


A numerical method is used to relax a braided magnetic field to an ideal force-free
equilibrium; that equilibrium is found to be smooth, with only large-scale current structures.
However, the equilibrium is shown to have a highly filamentary integrated parallel current
structure with extremely short length-scales.
An analytical model is developed to show that, in a coronal situation, the length scales associated
with the integrated parallel current structures will rapidly decrease with increasing
complexity, or degree of braiding, of the magnetic field.
Analysis shows the decrease in these length scales will, for any finite resistivity, eventually become 
inconsistent with the stability of a force-free field.
Thus the inevitable consequence of the magnetic braiding process is shown to be
a loss of equilibrium of the coronal field, probably via magnetic reconnection events.

%

\end{abstract}

{\bf Keywords:} MHD; magnetic fields; Sun: corona.

\newpage
\section{Introduction}
\label{sec:intro}

The heating of the solar corona as a result of the braiding of its constituent magnetic loops
is one of the fundamental theories proposed to explain coronal heating. Despite its long history 
relative to the subject area -- originating with 
the early notion of `topological dissipation'  (Parker, 1972) -- 
the braiding concept is highly controversial, with no general consensus reached as to the 
exact mechanism or even its general viability.

Coronal loops are typically modelled as consisting of ideal plasma threaded by a force-free
magnetic field.
With the ends of each loop anchored in the turbulent photosphere and Alfv{\`e}n travel times along 
a loop much faster than the time-scales associated with convection, the field evolution 
should be via a sequence of force-free states.
Parker (1972) argues that relaxation to a smooth equilibrium following an arbitrary perturbation
is impossible (excepting artificial cases with certain symmetries) and instead tangential discontinuities, 
corresponding to current sheets, must develop in the field.
The claim was first refuted in an analytical treatment by van Ballegooijen (1985)
who demonstrated the existence of smooth equilibria in non-symmetric 3D configurations
and under arbitrary continuous boundary perturbations.
He found singular current sheets appearing only with the application of discontinuous velocity 
fields at the boundary or in the case of a discontinuous magnetic field, corresponding to isolated 
magnetic flux patches in the photosphere (an idea followed-up in Priest {\it et al.} 2002).

Several other authors have also demonstrated classes of smooth equilibria
(Zweibel \& Li 1987, Longcope \& Strauss 1994, Craig \& Sneyd 2005)
and numerous authors have tackled the problem via numerical simulations 
(e.g.~Miki{\'c} {\it et al.} 1989;  Longcope \& Sudan 1994;  Galsgaard \& Nordlund 1996)
finding  equilibria that are non-singular -- although they do often contain current structures on a smaller
scale than that of the footpoint displacements.
However, it can be argued that numerical simulations may have an inherent resolution problem 
in allowing current sheet formation to be recognized.
In their analytical treatment  Craig \& Sneyd (2005) demonstrated the existence of smooth 
equilibria for any footpoint disturbance and for arbitrary compressibility.  Significantly, their 
argument was further advanced through the use of a Lagrangian relaxation scheme that 
enabled the stability of smooth solutions to be demonstrated, even for significant footpoint 
displacements.  

 
While no firm conclusion has been reached, either analytically or numerically,
on the capacity of magnetic braiding to produce current sheets in the corona,
it is clear  that the present thinking regarding magnetic braiding is fundamentally 
based on ideas developed from classical two-dimensional (2D)  reconnection theory 
(Sweet 1958; Parker 1957; Petschek 1964) namely that, in  order to allow for rapid reconnection, 
extremely small-scales in the electric current (compared with the large-scale 
magnetic field structure) are required.
In three dimensions  (3D)  a necessary condition for the occurrence of reconnection  is a non-zero integrated
 electric field parallel to the magnetic field (Hesse \& Schindler 1988, Schindler {\it et al.} 1988).  In particular, 
reconnection may take place in the absence of any topological features such as magnetic null 
points.   As such it has quite different characteristics, both in terms of onset and evolution, to the 
two-dimensional case.
For some examples of 3D reconnection in the absence of a null point, the situation
relevant to magnetic braiding see, for example, Hornig \& Priest (2003) and Pontin {\it et al.} (2005).
Hence the results of magnetic braiding should be reassessed with the proper 3D reconnection
criteria in mind.
In particular the development of parallel electric fields as a result of  braiding processes should be 
examined, together with their consequences. 
We begin to address these issues in this paper, within the framework of resistive MHD.

The outline of the paper is as follows.  We build an analytical model for a braided magnetic field
between two   parallel plates  (Section~\ref{subsec:construct}) and describe a numerical 
relaxation scheme in which we relax this field to a force-free state, as expected for the solar corona
(Section~\ref{subsec:nm}).  
We then analyse the result of the relaxation process in terms of both the current structure
and integrated parallel currents (Section~\ref{sec:results}) 
and discuss how these structures depend on the  degree of braiding of the field
(Section~\ref{sec:complex}).
Results are discussed in Section~\ref{subsec:imp} and conclusions drawn in Section~\ref{sec:conc}.

\section{Methods}
\label{sec:methods}

\begin{figure}
\begin{center}
\includegraphics[width=0.25\textwidth]{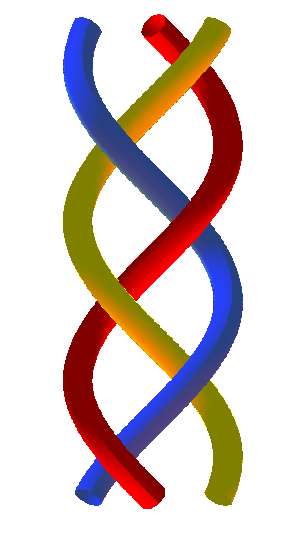}
\includegraphics[width=0.3\textwidth]{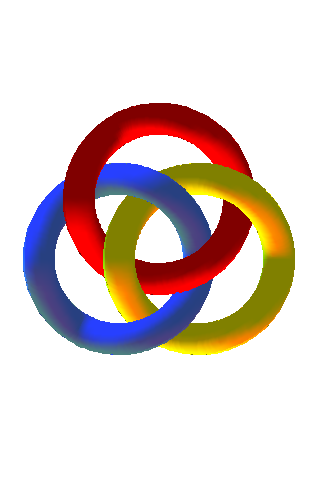}
\end{center}
\caption{
({\it left})  Illustration of  the pigtail braid, on which the braided magnetic field 
modelled here is based. The pigtail braid is the braid equivalent of the
Borromean rings  ({\it right})}
\label{fig:pigtailandborromean}
\end{figure}

In this paper we are interested in how  braided magnetic fields relax to a force-free equilibrium
and specifically in the nature of such equilibria.
We begin by describing a general method for constructing braided fields 
and then concentrate on one particular field structure, a configuration based on the 
Borromean rings.   
  This situation has been chosen since it represents a truly braided field with a number of
 advantageous solar-like properties (as described in Section~\ref{subsec:construct}), lends itself
 to analytical modelling  and, in addition, can be modified in a way to allow for easy comparison 
 with more and less complex fields.
 We do not concentrate on the details of how this particular field would be built up in the corona, rather 
 taking it as a fairly representative example of a complex field.
The construction method for this and similar fields is described in the following section.

\subsection{Braid construction}
\label{subsec:construct}

The braid considered here is modelled on a standard `pigtail' braid.  The pigtail braid consists of 
three strands, braided as shown in Figure~\ref{fig:pigtailandborromean} ({\it left}).
By way of motivation for this choice we suppose that
the braiding of field lines anchored in the photosphere
occurs in a random manner, that is, left-handed and right-handed twists occur with the same
probability. We therefore wish to take a braid which has no net twist.  More precisely, if we close the
braid to form a link, by identifying the upper and lower planes in 
Figure~\ref{fig:pigtailandborromean}, then the pairwise linking of strands should vanish.
The simplest link with this property is the Borromean rings 
(see right hand image of Figure~\ref{fig:pigtailandborromean}) which have the property 
that any two of the three rings are unlinked whilst the whole link itself is non-trivial, i.e.~cannot
be taken apart.  Our pigtail braid is the braid equivalent of the Borromean rings.
The vanishing net twist of our configuration is also the most conservative case.
It is easy to drive currents in a model by highly twisting the field, a situation
that is not observed in the corona,  whilst our choice represents an essentially
twist-free situation.
A further advantage of the braid   
 is that it can be considered as the concatenation of three
identical parts, each consisting of a pair twists (intersection of strands),  
one positive and one negative.  As such  it is easy to study the effect of increasing 
braid complexity through a concatenation of increasing numbers
of these elementary parts.

The magnetic fields representing our braids are constructed as follows.  Each field
 consists initially of a uniform background 
field ($b_{0} \hat{\bf{z}}$), superimposed on which are a number of isolated 
magnetic flux rings, corresponding to the crossings of strands in our braid.
The basic form of these flux rings is given, in cylindrical polar coordinates, by
\begin{equation}
\label{eq:fieldinpolars}
\mathbf{B}_{c} = 2 b_{0} k \frac{r}{a} \textrm{exp} \left( -\frac{r^{2}}{a^{2}} - \frac{z^{2}}{l^{2}} \right) \mathbf{e}_{\phi},
\end{equation}
with the parameters included allowing for variation in the field strength ($k, b_{0}$), as well as the
radius ($a$) and vertical extent ($l$) of the region of toroidal field.
The effect of adding the field given by equation~\eqref{eq:fieldinpolars} to the background field $b_{0} \hat{\bf{z}}$
is to create a region of non-uniform twist within the otherwise straight vertical field, with the maximum degree of twist controlled by 
the parameter $k$.  An  expression corresponding to equation~\eqref{eq:fieldinpolars}  may be found in Cartesian 
coordinates with the field centre transposed from $r=0, \ z=0$ to $x=x_{c}, \ y=y_{c}, \  z=z_{c}$.  We label this
field $\mathbf{B}_{c}$:
\begin{equation}
\label{eq:unittwist}
\mathbf{B}_{c} = 
2  \frac{b_{0} k}{a}  \textrm{exp}\left(
\frac{ \scriptstyle-\left(x-x_{c}\right)^{2}-\left(y-y_{c}\right)^{2}}
{ \scriptstyle a^{2}} - \frac{ \scriptstyle \left(z-z_{c}\right)^{2}}{\scriptstyle l^{2}}  \right)
\left( -  \left(y - y_{c} \right) \mathbf{\hat{x}}  + \left(x - x_{c} \right) \mathbf{\hat{y}}  \right),
\end{equation}
and let  $\mathbf{B_{c_{i}}}$ denote the field $\mathbf{B}_{c}$ taken with the parameters 
$ {\bm{c}}_{i} = \left( x_{c,i},  \ y_{c,i}, \ z_{c,i}, \  k_{i}, a_{i}, l_{i} \right)$.
Braided fields of various complexity may then constructed as
\begin{equation}
\label{eq:various}
\mathbf{B} = b_{0} \hat{\bf{z}} + \sum_{i=1}^{n}\mathbf{B_{c_{i}}}, 
\end{equation}
with appropriately chosen parameter sets $ {\bm{c}}_{i}$  ($i=1 \ldots n$).  In this paper we set $b_{0}=1$ throughout.

The braided fields constructed in this manner are taken as an initial condition in  numerical relaxation experiments.
An alternative would have been to begin with a uniform magnetic field between lower and upper boundaries
and use a numerical technique to simulate photospheric motions  and obtain a braided field and  to then proceed
to relaxation techniques.
However, here our primary interest lies in the relaxation process and its consequences and so 
we take the analytical models  as a proxy for a field constructed through simulated photospheric motions
The advantages of this approach lie in a significant saving in computational time,
the ability to easily adopt  fields with the aforementioned braid properties 
and a model that lends itself to analytical analysis.

\begin{figure}
\centering
\begin{tabular}{ll}
\multirow{2}{0.25\textwidth}[7.01cm]{
\subfigure[Some particular field lines within $E^{3}$.]{\label{fig:edge-3na}\includegraphics[width=0.165\textwidth]{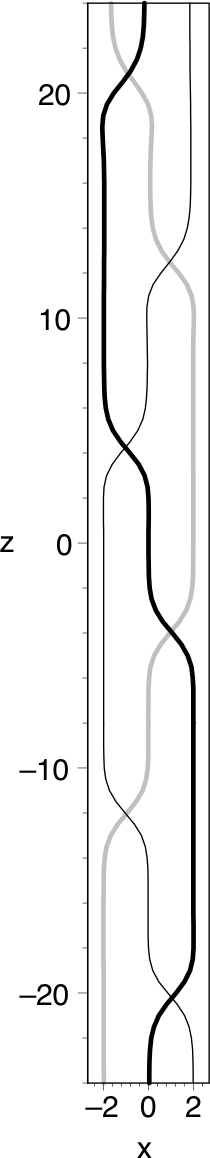}}
} &
\subfigure[Illustrating the same field lines as in (a) with a different aspect 
ratio and view-point to show more clearly the braided nature of the field.]
{\label{fig:edge-3nb} \includegraphics[width=0.4\textwidth]{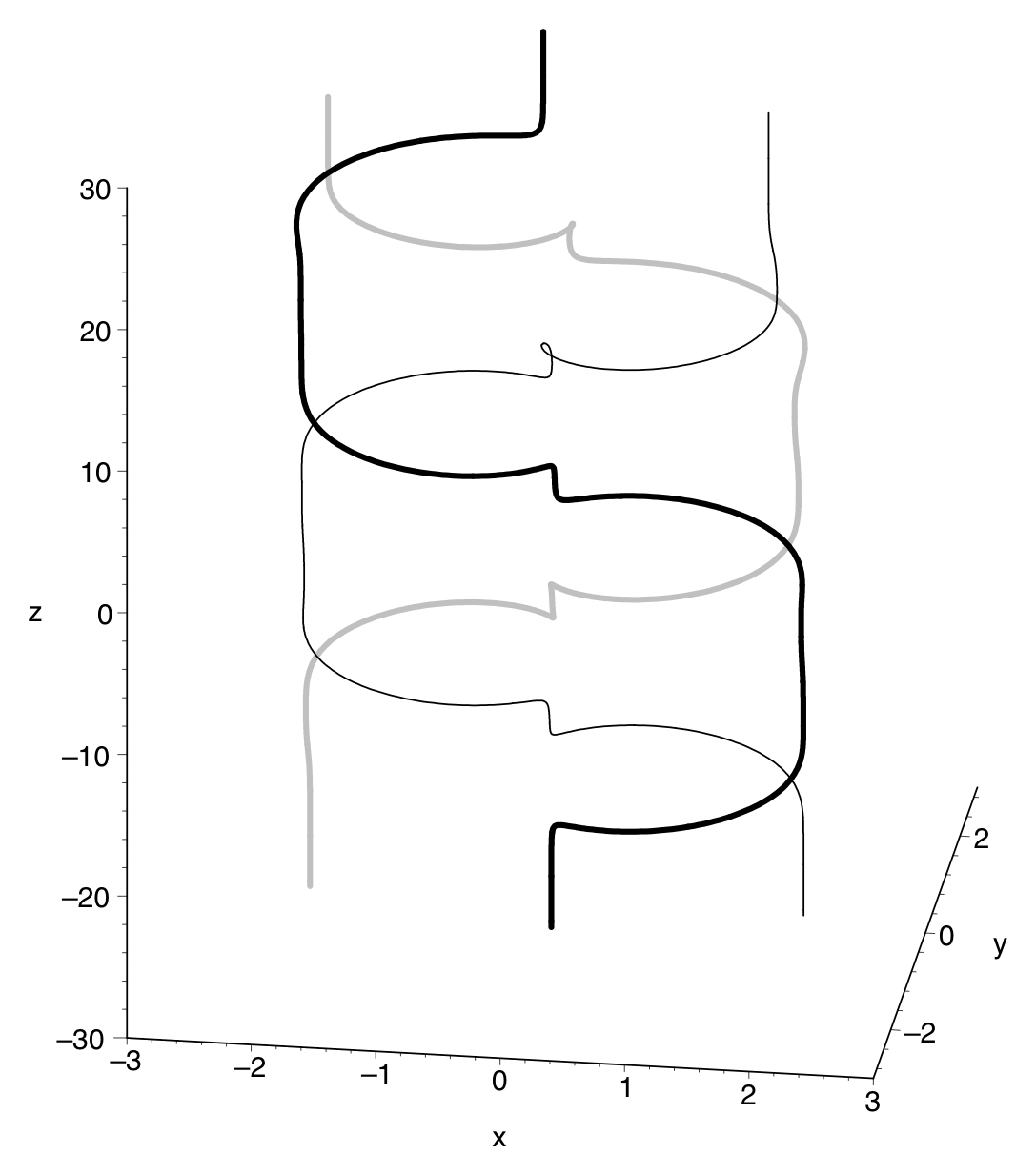}}               \\[4ex]
 & 
 \subfigure[Isosurfaces of current, $j = j_{max}/4$.]{\label{fig:edge-3nc}\includegraphics[width=0.4\textwidth]{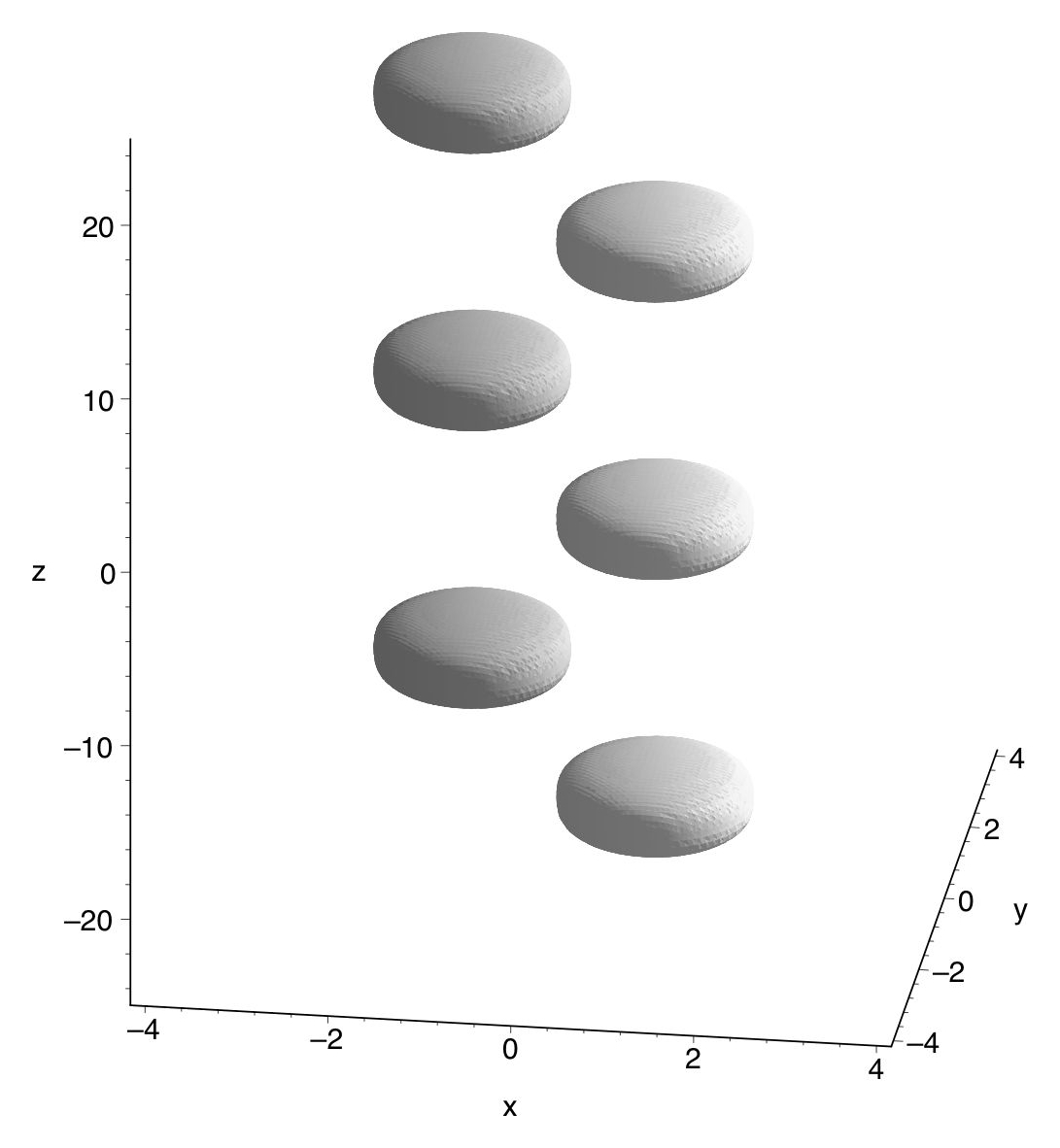}}                \\
 \end{tabular}
\label{fig:3lines}
\caption{Initial state for the braided magnetic field $E^{3}$.}
\end{figure}

To construct the pigtail braid on which we base the majority of our analysis 
(Sections~\ref{sec:methods} and \ref{sec:results}) we take  the parameter sets
${\bm c_{1}}= \left(1,0,-20, 1,\sqrt{2}, 2 \right)$,  ${\bm c_{2}}= \left(-1, 0,-12, -1, \sqrt{2},  2 \right)$,
${\bm c_{3}}= \left(1,0, -4, 1, \sqrt{2},  2 \right)$, ${\bm c_{4}}= \left(-1, 0, 4,  -1, \sqrt{2}, 2 \right)$,
${\bm c_{5}}= \left(1,0, 12,1,\sqrt{2}, 2 \right)$, ${\bm c_{6}}= \left(-1, 0, 20,-1, \sqrt{2}, 2 \right),$ 
and let 
\begin{equation}
\label{eq:6braid}
\mathbf{B} =  b_{0} \hat{\bf{z}} + \sum_{i=1}^{6} \mathbf{B_{c_{i}} }.
\end{equation}
As previously mentioned the field is the concatenation of three identical parts.  The
`elementary' unit, consisting of one positive and one negative twist, we label $E$ for
convenience and so refer to the pigtail braid as $E^{3}$.
In a similar manner we may construct the fields $E^{2}$, $E^{4}$ etc.
The construction of $E^{3}$ is such that the field lines threading the regions of maximum twist form a Borromean
braid as shown in Figure~\ref{fig:3lines}.  However only a small proportion of the total flux is braided in this manner;
field lines typically pass through fewer of the regions of toroidal field, turning by a smaller angle as they do so.
The regions of twisted field result in six isolated regions of  current, as shown in Figure~\ref{fig:3lines}(c).

The field is defined in infinite space, while to realistically model coronal loops we should take lower and 
upper boundaries, representing the photosphere, to which the field is perpendicular.
For the analytical work that follows, the lower boundary can be conveniently considered by taking the
limit $z \rightarrow -\infty$ and, similarly, the upper boundary as $z \rightarrow \infty$.
Practically however, due to the exponential decay of the toroidal 
field components, a short distance above and below the regions of toroidal field centered at
$z= \pm 20$, the field can be approximated as vertical.  
Accordingly, for the numerical investigations we define the domain as 
$x \in [-4,4], y \in [-4,4], z \in [-24,24]$.  Taking this domain it is worthwhile to note that the difference in magnetic
energy between $E^{3}$, as defined by Equation~\eqref{eq:6braid}, and the potential field satisfying the same
boundary conditions (i.e.~the uniform field $b_{0} \hat{\mathbf{z}}$) is just 3.08\%;  the 
field is close to potential in this respect.
This is in accordance with observations of coronal fields (Solanki {\it et al.} 2006).

We now proceed to describe the numerical method we will use to relax $E^{3}$ to a force-free state
whilst keeping the field on the boundaries fixed.

 \subsection{Numerical methods}
  \label{subsec:nm}
  
Whilst, as previously mentioned, the field in the solar corona is expected to be  largely force-free, 
i.e. in an equilibrium state satisfying
 \begin{math}
\left( \mathbf{\nabla} \times \mathbf{B} \right) \times \mathbf{B} = 0,
\end{math}
 the initial state for $E^{3}$, as given by Equation~\eqref{eq:6braid}  is not in such an equilibrium. 
We therefore seek to relax that field towards a force-free equilibrium 
while preserving the topology of the field, i.e.~the relaxation must correspond to an ideal 
plasma evolution within the volume and have vanishing velocity on the boundaries.
 For this we 
 use the  3D Lagrangian magneto-frictional ideal relaxation scheme described in Craig \& Sneyd (1986).
 The advantage of a Lagrangian scheme is that since the numerical grid and, due to the ideal evolution, 
 the magnetic field lines,  moves with the plasma elements,  grid points accumulate in 
regions where length scales become small. This  enables a better resolution of these small-scale features 
than would be achieved with a similar Eulerian code and the same number of grid points.
 
Details of the code can be found in Craig \& Sneyd (1986, 1990) and are, therefore, only briefly summarised  here for convenience.  
Since the relaxation is ideal, both the line element $\delta \mathbf{x}$  that joins two fluid 
particles and the vector $\mathbf{B}/\rho$ must follow the same time evolution, that is
\begin{displaymath}
\frac{D}{Dt} \left( \frac{\mathbf{B}}{\rho} \right) = \left( \frac{\mathbf{B}}{\rho} \cdot \mathbf{\nabla} \right) \ \mathbf{v},
\end{displaymath}
and similarly for $\delta \mathbf{x}$;
the code makes use of this by invoking the same Lagrangian description for both. 
In order to realize the equilibria, a fictitious equation of motion  that
guarantees a monotonic decrease in the magnetic energy of the system is adopted.
The momentum equation takes the form
\begin{displaymath}
\nu \mathbf{v} = \mathbf{j} \times \mathbf{B},
\end{displaymath}
i.e.~a frictional term proportional to the velocity of the fluid, $\mathbf{v}$,
is taken ($\nu \sim 1$) and inertial effects ignored.
The choice is also advantageous in that it is a parabolic system, so
the possibility of shock waves, and is justified since our primary interest is in the final 
stable equilibrium rather than the path to it.     Additionally, 
this choice of equation has been shown not to alter the stability or state of the final equilibrium.
Magnetic flux and $\mathbf{\nabla} \cdot \mathbf{B}$ are both conserved by the relaxation.
The numerical technique itself is an implicit (ADI) unconditionally stable scheme.

\section{Force-Free State}
\label{sec:results}


We now take the initial state for $E^{3}$ (as given by Equation~\ref{eq:6braid}) and use the 
numerical method described  in Section~\ref{subsec:nm}
to relax it towards a force-free equilibrium, with the field on the upper and lower boundaries 
($z=\pm 24$)  held fixed by line-tying at those boundaries.
The experiment has been run at various resolutions with the results presented here 
being from a $161^{3}$ computational grid  and a relaxed state with
$\left(\mathbf{j} \times \mathbf{B}\right)_{\max} \approx 2 \times 10^{-2}$ 
(a figure which may be compared with the equivalent initial value of $1.38$).
Numerical accuracy issues prevent further relaxation but
we view this state as a good approximation to force-free;
for a discussion 
see Section~\ref{sec:conc}.


\begin{figure}[h]
\begin{center}
\subfigure[]{\label{fig:edge-3e}\includegraphics[width=0.5\textwidth]{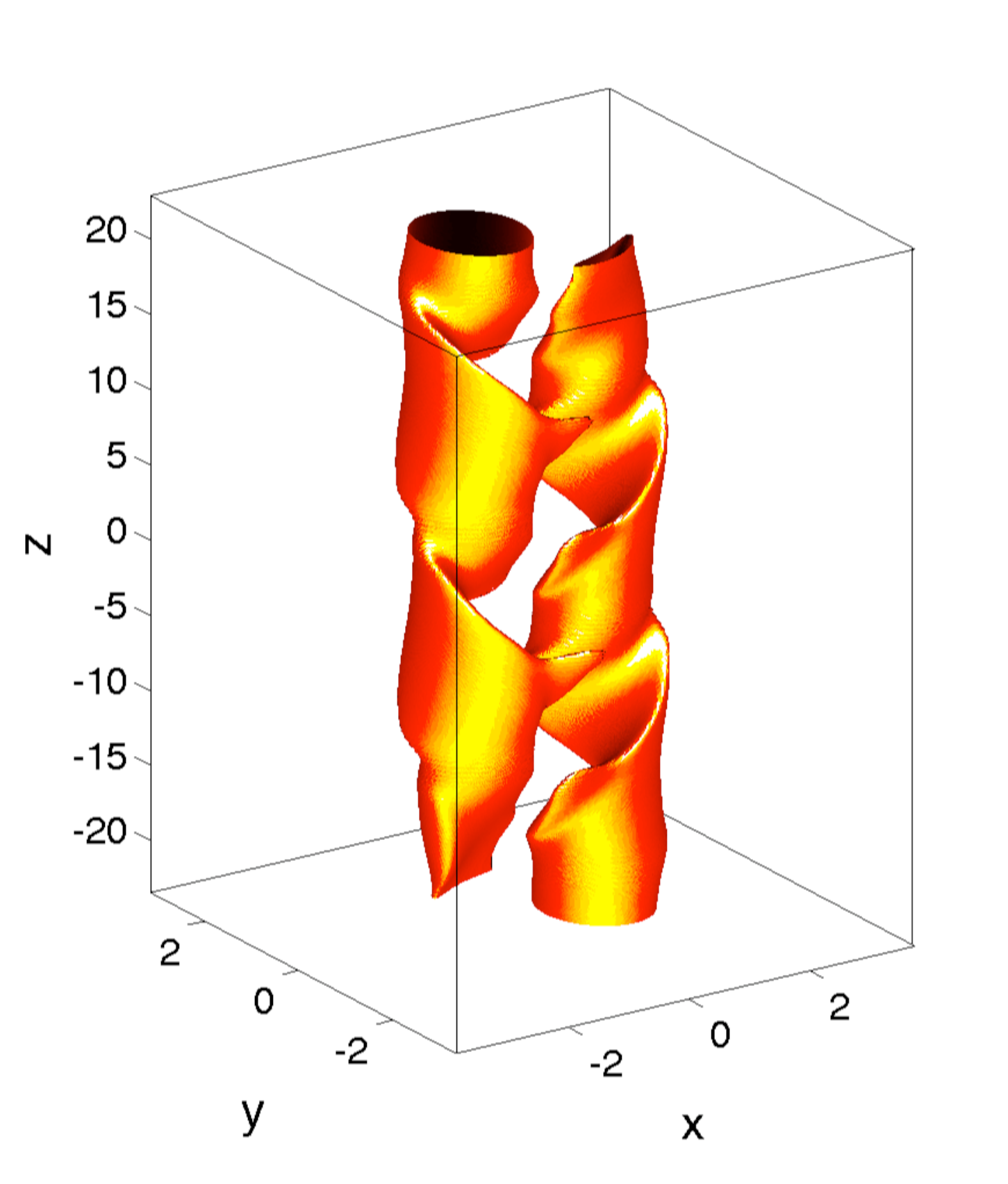}}
\subfigure[]{\label{fig:edge-3f}\includegraphics[width=0.28\textwidth]{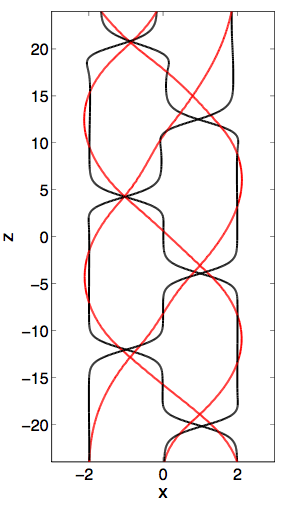}}
\end{center}
\caption{
(a) Isosurfaces of current density, ($\vert j \vert = 0.45$) in the relaxed equilibrium. 
(b) Particular field lines in $E^{3}$ in the initial state (black) and the equilibrium state (red). }
\label{fig:relaxed}
\end{figure}

The final equilibrium is found to be smooth in accordance with the predictions of van Ballegooijen (1985)
and others, for example, Craig \& Sneyd (2005).
 Figure~\ref{fig:relaxed}(a) shows an isosurface of current, $\left| j \right|$,  in the
force-free state.  Two spiraling current structures are seen to extend vertically throughout the domain.
No current sheets (singular or non-singular) are formed in the relaxation, indeed whilst initially the maximum absolute 
value of the  electric current is $\vert j \vert_{max} \sim 2.83$, in the final state this value is 
$\vert j \vert_{max} \sim 1.47$.  
Considering the field structure itself, some field lines illustrative of $E^{3}$ in both the initial
and final states are shown in Figure~\ref{fig:relaxed}(b). The same three field lines are plotted in 
both cases -- the boundary conditions allow us to identify these exactly.
 As illustrated in the figure, the twist in the  initial configuration has become distributed evenly along field lines
 although they are, of course, still braided since the relaxation was ideal.
 The magnetic energy in the relaxed, equilibrium state is just 0.96\% in excess of potential,
a figure which may be compared with the 3.08\% excess of the initial state (as 
 discussed in  Section~\ref {subsec:construct}).

That a smooth force-free equilibrium has been found for $E^{3}$ 
adds weight to the body of evidence against the Parker hypothesis (Parker, 1972) and 
lends support to the  arguments of van Balleooijen (1985) and Craig and Syned (2005)
that smooth coronal equilibria may be found after arbitrary photospheric evolutions.
However, as discussed in Section~\ref{sec:intro}, in three dimensions it is the integrated 
electric field component parallel to the magnetic field that is the crucial quantity for reconnection.
Since within the framework of resistive MHD, as considered here, Ohm's law is  given by 
\begin{math} \mathbf{E} + \mathbf{v} \times \mathbf{B} = \eta \mathbf{j} \end{math},
the parallel electric fields are directly related to parallel electric currents by the relation  
\begin{math}
\int E_{\parallel} dl = \int \eta j_{\parallel} dl = \eta \int j_{\parallel} dl,
\end{math}
where the last equality holds in the case of uniform resistivity $\eta$  (in practice, $\eta$ may be spatially 
varying, dependent, for example, on the current density itself).
These considerations motivate us to consider  the nature of $\int j_{\parallel} dl$. Results are
detailed in this section and their implications for the corona discussed in Section~\ref{subsec:imp}.


\begin{figure}
\begin{center}
\includegraphics[width=0.9\textwidth]{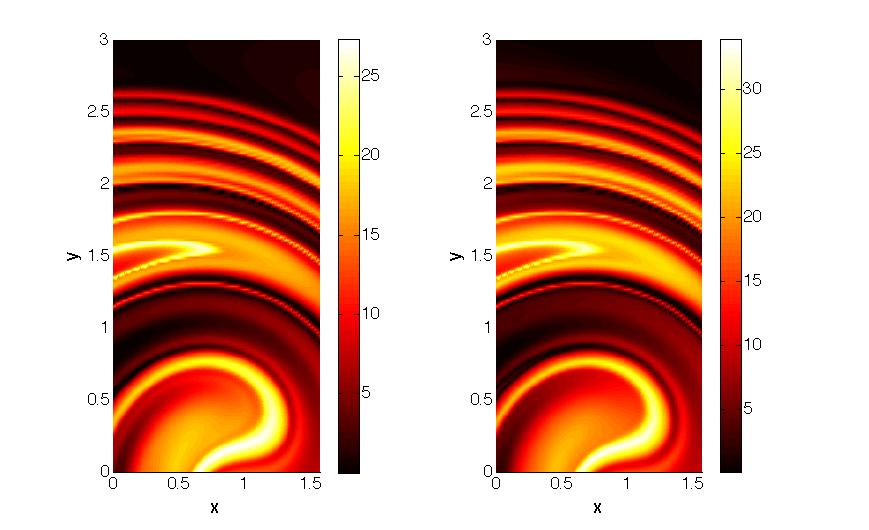}
\end{center}
\caption{ Contours of $\left| \int{j_{\parallel} \textrm{d}l } \right| $  for $E^{3}$
in the initial ({\it left}), and equilibrium ({\it right}), states.  The integral is taken along field-lines starting at the given location
on the lower boundary through to the upper boundary.}
\label{fig:jparcf}
\end{figure}

We calculate the integrated parallel currents,  $\int j_{\parallel} dl$, along field lines in $E^{3}$, 
both for the equilibrium state and, for  purposes of comparison,  the initial state.
 Specifically,  for a large number of points on the lower boundary of the domain ($z=-24$) we integrate
the parallel current along the field line starting from that location until it reaches the upper boundary ($z=+24$). 
Figure~\ref{fig:jparcf}  shows a contour plot of the result for  a subsection of the domain. 
In the figure the contour value indicates the total absolute  value of integrated parallel current,
\begin{math}
\left|  \int j_{\parallel} \textrm{d}l  \right|,
\end{math}
for the field line starting at that given location.
Two important  points about the structure of the integrated parallel currents are illustrated:
\begin{enumerate}
\item
The width of the layers of enhanced  $ \left| \int j_{\parallel} \textrm{d}l \right|$
is very small compared with both the size of the domain and the typical scales of $\mathbf{j}$ and $\mathbf{B}$.
This statement is made more precise in the following text.
\item
The structure, or distribution, of $ \left| \int j_{\parallel} \textrm{d}l \right|$
is approximately preserved in the relaxation process, although its peak value increases slightly
(from $27.04$ to $33.86$).
\end{enumerate}

The second of these points, the near preservation of
 $ \left|  \int j_{\parallel} \textrm{d}l  \right|$ along field lines during the relaxation
 process, has the consequence that the key properties of the equilibrium field
 may be deduced from the initial, non-equilibrium state.
 This allows us to examine  the distribution of  
$ \left|  \int j_{\parallel} \textrm{d}l  \right|$ in various braided field configurations
using only the initial state.
 This state is imposed as a (physically motivated) analytical dependence
and so we may quickly and accurately determine the 
distribution on the lower boundary of integrated parallel currents along field lines
without resorting first to computationally expensive relaxation experiments.

\begin{figure}
\includegraphics[width=\textwidth]{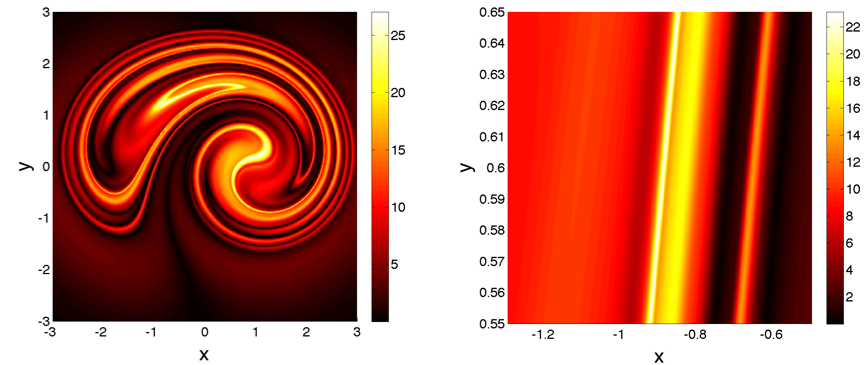}
\caption{ Contours indicating the value of $\left| \int{j_{\parallel} \textrm{d}l } \right| $ 
along field-lines in the initial state of $E^{3}$ for ({\it left}) the full domain and ({\it right})
 a small subsection only.}
\label{fig:jpar6braid}
\end{figure}

We focus  now on the second of our key points, the width of the integrated parallel currents.
Figure~\ref{fig:jpar6braid} gives contours of that quantity in the initial state (comparable to 
the equilibrium state as discussed above) for both the full domain and for a small subsection of it.
The integrated parallel currents are typically concentrated in thin layer-like structures,
the exception being the two broader peaks evident in the left hand image of 
Figure~\ref{fig:jpar6braid}
The right hand image of this figure shows one such typical layer in some detail.  
This layer has a half-width at half-maximum of  $0.0167$.
The width be compared with the typical scales of the electric current, $\mathbf{j}$, in the equilibrium
-- in the horizontal direction the  half-width at half-maximum of the current layer is $\sim 0.74$; 
the typical scale of the integrated parallel currents is $3\%$  of this value.
We address the possible consequences of the existence of these small scales
in the integrated parallel current in Section~\ref{subsec:imp}.

Recall that  in resistive MHD the parallel electric fields and parallel electric currents
are related via 
\begin{math}
\int E_{\parallel} dl = \int \eta j_{\parallel} dl, 
\end{math}
and that, in the case of uniform resistivity,
$\int j_{\parallel} dl$ (as considered here) will give us the structure of $\int E_{\parallel} dl$.

In practice, the resistivity, $\eta$, may not be spatially uniform but dependent, 
for example, on the current  density, $\mathbf{j}$.  
In the force-free equilibrium found we have $\left| j \right| \sim \left| B \right|$
and, in addition, $\left| B \right|$ is dominated by the constant $B_{z}$ component.
Hence the current  here is essentially constant along field lines.  
In the perhaps more realistic case of a 
 non-uniform resistivity,   a dependence such as $\eta \sim \vert j \vert ^{n}$ ($n>0$), for 
 example,
would therefore lead to integrals like $\int j^{n+1} dl$.  These  would show even
more enhanced structures and so taking $\eta$ as constant we are  considering
the most conservative case.

We now proceed to consider how changing the
 the complexity of the braided field affects results.

\section{Effect of Braid Complexity}
\label{sec:complex}

The initial state in $E^{3}$, as given by Equation \ref{eq:6braid}, consists of six 
isolated magnetic flux rings  (given in that expression by 
$\mathbf{B_{c_{i}}} $, $i=1 \ldots 6$)  superimposed onto a uniform 
background field ($b_{0} \hat{\mathbf{z}}$), the effect being to create six regions of isolated twist
within that otherwise straight field.
As discussed in Section~\ref{subsec:construct}
fields similar to $E^{3}$ may also be defined, using Equation \eqref{eq:various} and the appropriate 
choice of parameters (guided by Equation \ref{eq:6braid}) to maintain the same essential pattern in the braid, 
including the alternating sign of twist with height, but with varying numbers of twisted field regions.    
More precisely, various fields $E^{i}$ ($i = 1,2,3, \ldots$) may be constructed by reference to the
`elementary' unit, $E$, with magnetic field given by 
 \begin{displaymath}
\mathbf{B} = b_{0} \hat{\bf{z}} + \sum_{i=1}^{2} \mathbf{B_{\hat{c}_{i}}}, 
\end{displaymath}
where 
${\bm {\hat{c}_{1}}}= \left(1,0,-4, 1,\sqrt{2}, 2 \right)$ and  
${\bm { \hat{c}_{2}}}= \left(-1, 0,4, -1, \sqrt{2},  2 \right)$,
and choosing a suitable composition.
We name these fields according to the number of concatenations taken, so $E^{2}$ consists
of four twisted regions, $E^{4}$ of eight, etc.


We now analyse the pattern of integrated parallel currents along field lines for a number
of fields $E^{i}$ 
in a similar manner to that carried out for $E^{3}$.
The aim is to deduce how the width of the $ \left| \int j_{\parallel} \textrm{d}l \right|$
structures varies with braid complexity.  Our field construction method allows the complexity
of braided field $E^{n}$ to be denoted by $n$; we take an increase in $n$ to be generally representative
 of an  increase in the degree of braiding of the coronal field due to additional photospheric motions.

\begin{figure}
\begin{center}
\subfigure[]{\label{fig:edge-4a}\includegraphics[width=0.42\textwidth]{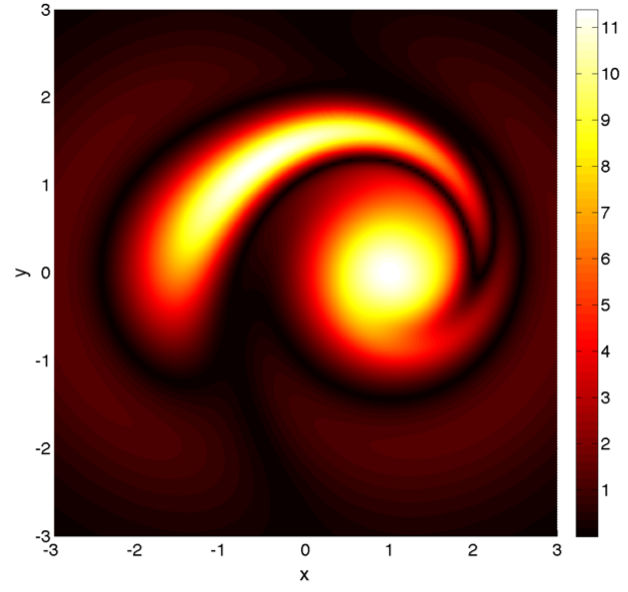}}
\subfigure[]{\label{fig:edge-4b}\includegraphics[width=0.48\textwidth]{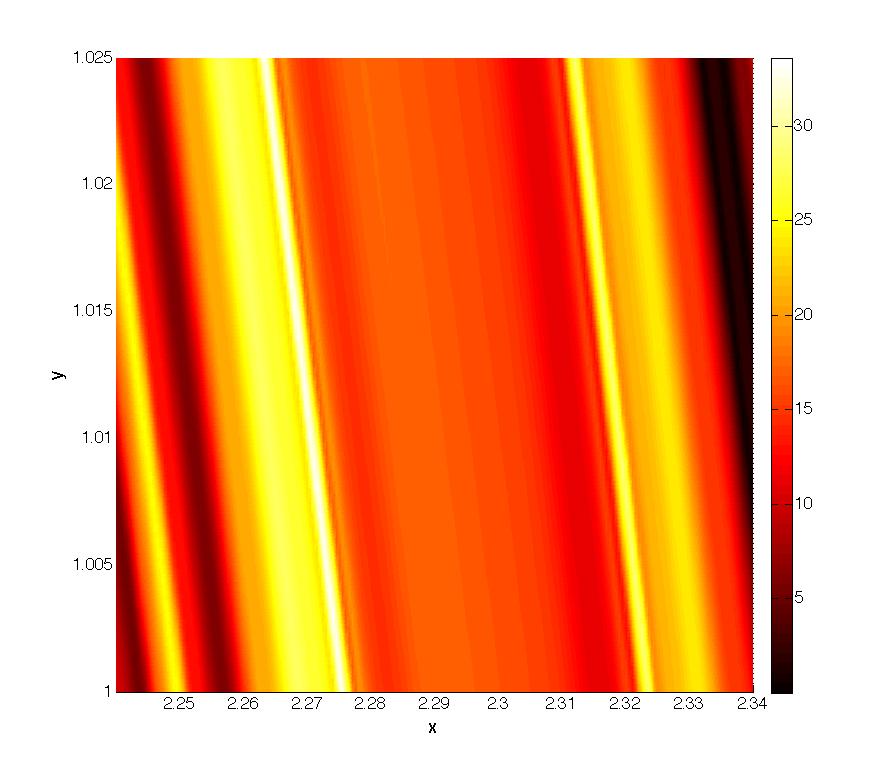}}
\end{center}
\caption{ Contours of $\left| \int{j_{\parallel} \textrm{d}l} \right|$ along field lines for 
appropriate portions of the domain in (a) $E$, and (b) $E^{5}$.}
\label{fig:2braid}
\end{figure}

Figure~\ref{fig:2braid} shows contours of $\left| \int{j_{\parallel} \textrm{d}l } \right|$  
for portions of the domain for both $E$ and $E^{5}$.  The scale of the domain shown in the
pictures has been chosen to allow the length-scales of $\left| \int{j_{\parallel} \textrm{d}l } \right|$
to be clearly seen.
For $E$ the maximum value of  $\left| \int{j_{\parallel} \textrm{d}l } \right|$ is lower ($\left| \int{j_{\parallel} \textrm{d}l } \right|_{\max}=11.38$)
Additionally,  the characteristic widths of the distribution are much broader.
For $E^{5}$ the  peaks in $\left| \int{j_{\parallel} \textrm{d}l } \right|$ 
are much narrower (as evidenced by the scale of the domain shown) and have a higher value
($\left| \int{j_{\parallel} \textrm{d}l } \right|_{\max}=33.62$
  in the section of the domain shown which has been arbitrarily chosen and, as such, is unlikely to encompass
the global maximum in $\left| \int{j_{\parallel} \textrm{d}l } \right|$).

 \begin{figure}
\begin{center}
\includegraphics[width=0.6\textwidth]{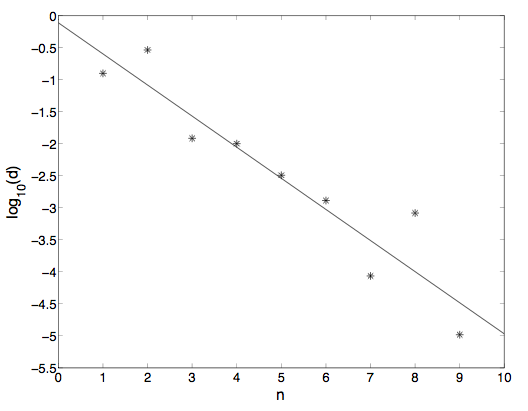}
\end{center}
\caption{Scatter plot showing the variation in the half-width at half-maximum ($d$) of the peak 
in $\left| \int{j_{\parallel} \textrm{d}l } \right|$
along the line $y=x$ ($0 \leq x \leq 2$) with $n$, where $n$ denotes the field $E^{n}$.
The  width of the integrated parallel current structures is seen to decrease exponentially with braid complexity --
the line is given by $d=\kappa 10^{-n/2}$, where $\kappa=0.77$.}
\label{fig:jpn}
\end{figure}

From these simple considerations we expect that as the complexity of a braided field increases
then the maximum in $\left| \int{j_{\parallel} \textrm{d}l } \right|$   increases while
 the width of the $\left| \int{j_{\parallel} \textrm{d}l } \right|$  structures, $d$, decreases.
 
 In order to determine the functional dependency $d(n)$ and the maximum value of
 $\left| \int{j_{\parallel} \textrm{d}l } \right|$
 with $n$ (number of elementary braids, $E$) we would have to make
 two-dimensional contour plots such as those in  Figure~\ref{fig:2braid}  for many higher
 numbers, $n$.
 Due to the increasing resolution required for these plots such a method is numerically
 very costly.
Instead we choose to consider integrated parallel currents along field lines originating along
the line $y=x$, $0 \leq x \leq 2$ on the lower boundary.
Taking just a 1D line cut has the disadvantages that the line may not be (indeed is unlikely to be) 
perpendicular to  $\left| \int{j_{\parallel} \textrm{d}l } \right|$ structures and may not cross regions of 
maximal $\left| \int{j_{\parallel} \textrm{d}l } \right|$.  
However the method might still be sufficient to allow a rough functional dependence of $d(n)$ 
to be determined.  For a number $E^{n}$ we consider
the half-width at half-maximum, $d$, of the maximum peak in the integrated parallel current value 
along this line.
We note that the value obtained may not be representative of the thinnest structures within the full domain
but the method should suffice in giving us an indication of the nature of the variation of $d(n)$.

The result is shown as a scatter plot in Figure~\ref{fig:jpn}; we deduce that the width of the integrated parallel current 
structures decreases exponentially with  increasing braid complexity --
the line superimposed is a fit to  the data with equation $d=\kappa 10^{-n/2}$, where $\kappa=0.77$.
Although this equation should not be taken to given any exact dependence of $d$ with $n$,
it gives a rough guide as to what we may expect.
In addition, we note that the distance between adjacent peaks in 
$\left| \int{j_{\parallel} \textrm{d}l } \right|$
also becomes smaller with increasing $n$.

The above described dependence of $d$ with $n$ might be expected from the 
following theoretical consideration.
If we assume that the maximum value of $\left| \int{j_{\parallel} \textrm{d}l } \right|$ in $E$ coincides
with the location where the field lines are stretched the most then we would expect
$d \sim d_{0}/\lambda$ where $\lambda$ is the 
eigenvalue of the field line mapping and $d_{0}$ the length-scale of the current distribution.
Subsequently an iteration of this map, as  appropriate for a concatenation of $E$ to form various braided fields
$E^{n}$, would result in the width $d$ of  $\left| \int{j_{\parallel} \textrm{d}l } \right|$ layers in $E^{n}$
of  $d \sim d_{0}/\lambda^{n}$.  This would occur provided there are sufficiently
many regions of maximal field line stretching 
that it becomes likely that a region with this property is mapped onto 
a similar region in the appended braid.
By the same argument we would expect the maximum value of  $\left| \int{j_{\parallel} \textrm{d}l } \right|$
to increase linearly with $n$.

The implications of the presence of narrow $\left| \int{j_{\parallel} \textrm{d}l } \right|$ layers within
a magnetic field structure are discussed in the following section.

 \section{Implications for Coronal Evolution}  
\label{subsec:imp}

Our results show clearly
that small-scale electric current structures are not a necessary requirement for
small-scale integrated parallel current structures.  The consequences of such a 
situation are less clear.
Whilst is is well known that in two dimensions thin current sheets can lead to rapid magnetic reconnection,
an equivalent condition for the three-dimensional case has not been deduced as yet.
In this section we show that small scales in the integrated parallel currents are a likely candidate; they are,
at least, inconsistent with the presence of a stable force-free non-equilibrium.


We used a numerical code to relax $E^{3}$ towards a force-free equilibrium state, $\mathbf{j} \times \mathbf{B} = 0$.
In such an equilibrium
the ideal Ohm's law, $\mathbf{E} + \mathbf{v} \times \mathbf{B} = \mathbf{0}$, may be satisfied
with $\mathbf{v}=\mathbf{0}$ and $\phi = 0 $, where $\mathbf{E} = - \mathbf{\nabla} \phi$ due to the
stationarity of the state,  a stable force-free ideal equilibrium.
In a real resistive plasma, such as the solar corona, a very small but finite resistivity, $\eta$, is
present and Ohm's law may be given by
\begin{equation}
\label{eq:ohm}
\mathbf{E} + \mathbf{v} \times \mathbf{B} = \eta \mathbf{j}.
\end{equation}
%
Then, since the equilibrium is time independent, $\mathbf{E} = - \mathbf{\nabla} \phi$ and the electric
potential $\phi$ may be deduced by integrating along  field lines. For a spatially uniform $\eta$, $\phi$
is given by
\begin{displaymath}
\phi = -\int \eta j_{\parallel} \ \textrm{d} l = -\eta \int j_{\parallel} \ \textrm{d} l. 
\end{displaymath}
The plasma velocity $\mathbf{v}$ then follows as
\begin{displaymath}
\mathbf{v} = \frac{\left( - \mathbf{\nabla} \phi - \eta \mathbf{j}\right) \times \mathbf{B}} 
{B^{2}}
= \frac{\left( - \mathbf{\nabla} \phi \right) \times \mathbf{B}} 
{B^{2}},
\end{displaymath}
where the second equality  holds  since the equilibrium field is force-free.
From this we estimate the plasma velocity as
\begin{displaymath}
\mathcal{O}\left(v\right) \sim \frac{\phi}{d}\frac{1}{B}, 
\end{displaymath}
since the narrow width $d$ of the parallel current structures will induce the
largest gradients in the electric potential.
Similarly, making an estimate of the size of the electric potential, 
\begin{displaymath}
\mathcal{O}(\phi)  \sim  \eta   \left|  \int j_{\parallel} \ \textrm{d} l   \right|  \sim \eta \frac{B}{\mu_{0} l} L,
\end{displaymath}
where $l$ denotes the width of the electric current structures
and $L$ the length over which the integration should be  carried out. 
Combining these estimates, and comparing the plasma velocity to the Alfv{\'e}n velocity, we find
\begin{displaymath}
\mathcal{O} \left(  \frac{v}{v_{A}}  \right)  
\sim  \frac{\eta}{\mu_{0} l}  \frac{L}{d} \frac{1}{v_{A}} = \frac{1}{S} \frac{L}{d}
\end{displaymath}
where $S$ is the Lundquist number based on the length-scale, $l$, of the current structures.

Taking  $E^{n}$ as an example, and taking $L$ fixed
(since in a realistic situation the length of a coronal  loop should remain 
roughly constant with braiding),
an approximate relation for $d$ is $d \sim 0.77 \times 10^{-n/2}$ (as discussed in Section~\ref{sec:complex})
and, accordingly (taking $L \sim 48$, as for $E^{3}$)
\begin{displaymath}
\mathcal{O}\left(  \frac{v}{v_{A}}  \right)  \sim  \frac{1}{S} \ 70^{n/2}
\end{displaymath}
Therefore for any finite value of $S$ we can find a value $n$ such that $v \gg v_{A}$ and the $n$-braid
cannot be in force-free equilibrium.
In other words,  the ideal approximation ($S \rightarrow \infty$)
in which a force-free equilibrium may be found for arbitrary $n$ must eventually break down, 
even if the current  structures themselves are of a large-scale.


Within the more general theory of the magnetic braiding of the solar corona (where $L$ may be, say, roughly 
the length of a coronal loop and remain approximately fixed with increasing braiding), 
from these models we expect the width of
integrated parallel current layers, $d$, to decrease exponentially as the coronal field becomes more and more braided by
photospheric motions.  As we have shown,
 this increase in complexity need not cause a decrease in the scale of the current structures themselves, i.e.~thin current 
sheets may not necessarily form (van Ballegooijen, 1985).
However, after a finite amount of braiding $d$ will become so small that, no matter how small the resistivity,
the equilibrium cannot be maintained.
One possibility is that the loss of equilibrium will be via magnetic reconnection.  In three dimensions the
rate of reconnection is given 
by $\int E_{\parallel} dl$ where the integral is taken along the field line where the quantity 
has its maximum value.
In order to determine where and how magnetic reconnection is triggered we would need to
take a full 3D MHD simulation of the domain.  However, the small-scale structures found in
$\left| \int{j_{\parallel} \textrm{d}l } \right|$  will necessitate a very high spatial resolution
for any such accurate MHD simulation.  For sufficiently high $n$ and low $\eta$ this is
beyond the reach of currently available computing power.  Whether for moderate $n$ and
comparatively high $\eta$ effects due to the existence of small length-scales in
$\left| \int{j_{\parallel} \textrm{d}l } \right|$  are already evident 
is the subject of a current investigation.


\section{Discussion and Conclusions}
\label{sec:conc}

In this paper we have readdressed the magnetic braiding process (Parker, 1972)
in view of the consideration that in three dimensions it is the integrated parallel electric
fields along field lines that are important for reconnection. In resistive MHD these are related
 to the integrated parallel currents,  not just the current structure itself.
The key outcome of the investigation is a theory that as the coronal field is braided the
integrated parallel current along field lines will develop smaller and smaller 
structures.  As a result we 
conclude  that relaxation to a smooth equilibrium is not possible for
a magnetic field with an arbitrary degree of braiding;
magnetic braiding by photospheric motions will, at some point, lead to a lack of equilibrium
in the coronal loop. 

In Section~\ref{subsec:construct}  a particular magnetic field configuration is constructed,   
modelled on the Borromean rings.  We refer to this field as $E^{3}$.
The initial configuration for $E^{3}$ is taken as a magnetic field between two parallel plates 
containing six isolated twisted regions in an otherwise vertical field.
Each twisted region has only a small amount of twist and the overall twist  of the configuration is zero.
Thus the braid can be taken to model a (straightened-out) coronal loop.  Its aspect ratio is 
such that, together with the amount of twist, it would likely appear straight in coronal conditions.


A Lagrangian numerical scheme was taken to relax the field $E^{3}$ towards a force-free
equilibrium with the field on the boundaries held fixed.  In the relaxed state the maximum 
Lorentz force inside the volume is two  orders of magnitude lower, at $\sim 10^{-2}$, than 
that initially present.  With the technique presently employed, numerical difficulties 
prevent further asymptotic relaxation towards $\mathbf{j} \times \mathbf{B} = 0$.
This is believed to be a result of the implementation of second-order
differencing on a highly deformed grid and is the subject of a  current investigation.
Nevertheless, in the final state  the twist in the configuration is evenly distributed 
along field lines and the current,  initially in the form of six isolated closed current regions, 
is in two smooth, twisted, layers extending throughout the volume.
Accordingly,  our approximation to a force-free state is a reasonable one 
that allows us to deduce the main results.

The equilibrium state is described in Section~\ref{sec:results}.
There is no evidence for  small-scales developing in 
the current structure, lending support to those arguing against 
Parker (1972) -- for example van Ballegooijen (1985) and Craig \& Sneyd (2005) -- 
who claim smooth force-free equilibria are possible under arbitrary boundary 
perturbations.
However,
in three dimensions it is the presence of an integrated electric field component parallel 
to the  magnetic field that is crucial for magnetic reconnection.   In resistive MHD this is 
related to the integrated parallel current structure along field lines
($ \int{j_{\parallel} \textrm{d}l }$), rather than the current itself.
This consideration motivated us to consider the  nature of 
$ \int{j_{\parallel} \textrm{d}l }$
in the braided field $E^{3}$, both for the initial and relaxed states.

The variation in $\left| \int{j_{\parallel} \textrm{d}l } \right|$ between field lines originating on the
lower boundary of the domain is examined in Section~\ref{sec:results}.
The  $\left| \int{j_{\parallel} \textrm{d}l } \right|$ structure is found 
to be nearly the same in both cases, i.e. approximately conserved
in the relaxation process,  and there is seen to be significant variation 
in the value of this quantity between field lines that intersect the lower boundary of the domain only a small distance apart.
 We refer to this characteristic as `thin integrated parallel current layers'.
During the relaxation process the width of these layers is preserved, whilst the 
peak values in $\left| \int{j_{\parallel} \textrm{d}l } \right|$ are slightly enhanced,
so creating larger  gradients in $\left| \int{j_{\parallel} \textrm{d}l } \right|$
between the layers.


In Section~\ref{sec:complex} the variation in the width, $d$, of these 
integrated parallel current layers
with  complexity, or degree of braiding, of the magnetic field
is considered through an examination of the fields $E^{n}$ for various $n$.
Here each increase in $n$ can be thought of as being the result of 
additional braiding  of the coronal field via photospheric motions.
With increasing $n$, $d$  is found to decrease exponentially.
More generally then, the width of the integrated parallel current layers within 
a configuration is expected to  decrease with increasing braiding.

In Section~\ref{subsec:imp} arguments are given relating to the consequences of narrow
$\left| \int{j_{\parallel} \textrm{d}l } \right|$ layers on the possibility of finding a force-free
 equilibrium.   It is concluded that  once
$L/d \sim S$ (where $L$ represents the length of the braided field lines and $S$ the Lundquist number)
-- a state that is inevitable as the magnetic braiding by photospheric motions continues --
a smooth equilibrium state can no longer be attained.
The precise manifestation of this lack of equilibrium is yet to be determined
but we consider it is likely to be via magnetic reconnection events.


\newpage

\noindent
{\large {\bf Bibliography}}

\noindent
{Craig}, I.~J.~D. and {Sneyd}, A.~D.,   ApJ, 311, 451 (1986) \\
{Craig}, I.~J.~D. and {Sneyd}, A.~D.,  ApJ, 357, 653 (1990) \\
{Craig}, I.~J.~D. and {Sneyd}, A.~D., Sol. Phys., 232, 41(2005)  \\
{Galsgaard}, K. and {Nordlund}, A\&A,  JGR, 101, 13445 (1996) \\
 {Hesse}, M. and {Schindler}, K., JGR, 93, 5559 (1988) \\
 {Hornig}, G. and {Priest}, E.,  Phys. Plasmas, 10, 2712 (2003) \\
   {Longcope}, D.~W. and {Strauss}, H.~R., ApJ, 437, 851 (1994) \\
 {Longcope}, D.~W. and {Sudan}, R.~N., ApJ, 437, 491 (1994) \\
{Miki{\'c}}, Z. and {Schnack}, D.~D. and {van Hoven}, G., ApJ, 338, 1148 (1989) \\
{Parker}, E.~N., JGR, 62, 509 (1957) \\
{Parker}, E.~N., ApJ, 174, 499 (1972)  \\
{Petschek}, H.~E., in `The Physics of Solar Flares',  ed., {{Hess}, W.~N.}, p. 425 (1964)  \\
   {Pontin}, D.~I. and {Galsgaard}, K. and {Hornig}, G. and {Priest}, E.~R., Phys. Plasma, 12, 052307 (2005) \\
 {Priest}, E.~R. and {Heyvaerts}, J.~F. and {Title}, A.~M., ApJ, 576, 533 (2002) \\
{Schindler}, K. and {Hesse}, M. and {Birn}, J., JGR, 93, 5547 (1988) \\
{Solanki}, S.~K. and {Inhester}, B. and {Sch{\"u}ssler}, M., Rep. Prog.  Phys., 69, 563 (2006) \\
{Sweet}, P.~A., in `Electromagnetic Phenomena in Cosmical Physics,   ed. {{Lehnert}, B.}, p. 123, (1958) \\
 {Van Ballegooijen}, A.~A., ApJ, 298, 421 (1985) \\
{Zweibel}, E.~G. and {Li}, H.-S., ApJ, 312, 423 (1987)

\end{document}